\documentclass[aps,prl,reprint,superscriptaddress,longbibliography]{revtex4-2}
\usepackage{graphicx}
\usepackage{float}
\usepackage{blindtext}
\usepackage{physics}
\usepackage{gensymb}
\usepackage{xcolor}
\usepackage{comment}

\begin{document}

\title{Bose-Einstein condensate source on an optical-grating-based atom chip for quantum sensor applications}

\author{R. Calviac}
\affiliation{Laboratoire Collisions Agr\'egats R\'eactivit\'e, UMR 5589, FERMI, UT3, Universit\'e de Toulouse, CNRS, 118 Route de Narbonne, 31062 Toulouse CEDEX 09, France}
\affiliation{LAAS-CNRS, Universit\'e de Toulouse, CNRS, 7 Avenue du Colonel Roche, 31400 Toulouse, France}
\author{A. Rouxel}
\affiliation{LAAS-CNRS, Universit\'e de Toulouse, CNRS, 7 Avenue du Colonel Roche, 31400 Toulouse, France}
\author{S. Charlot}
\affiliation{LAAS-CNRS, Universit\'e de Toulouse, CNRS, 7 Avenue du Colonel Roche, 31400 Toulouse, France}
\author{D. Bourrier}
\affiliation{LAAS-CNRS, Universit\'e de Toulouse, CNRS, 7 Avenue du Colonel Roche, 31400 Toulouse, France}
\author{A. Arnoult}
\affiliation{LAAS-CNRS, Universit\'e de Toulouse, CNRS, 7 Avenue du Colonel Roche, 31400 Toulouse, France}
\author{A. Monmayrant}
\affiliation{LAAS-CNRS, Universit\'e de Toulouse, CNRS, 7 Avenue du Colonel Roche, 31400 Toulouse, France}
\author{O. Gauthier-Lafaye}
\affiliation{LAAS-CNRS, Universit\'e de Toulouse, CNRS, 7 Avenue du Colonel Roche, 31400 Toulouse, France}
\author{A. Gauguet}
\affiliation{Laboratoire Collisions Agr\'egats R\'eactivit\'e, UMR 5589, FERMI, UT3, Universit\'e de Toulouse, CNRS, 118 Route de Narbonne, 31062 Toulouse CEDEX 09, France}
\author{B. Allard}
\email{allard@irsamc.ups-tlse.fr}
\affiliation{Laboratoire Collisions Agr\'egats R\'eactivit\'e, UMR 5589, FERMI, UT3, Universit\'e de Toulouse, CNRS, 118 Route de Narbonne, 31062 Toulouse CEDEX 09, France}

\date{\today}

\begin{abstract}
We report the preparation of Bose-Einstein condensates (BECs) by integrating laser cooling with a grating magneto-optical trap (GMOT) and forced evaporation in a magnetic trap on a single chip.
This new approach allowed us to reach Bose-Einstein condensate threshold with $6\times 10^4$ atom Bose-Einstein condensate of rubidium-87 atoms with a single laser cooling beam.
Our results represent a significant advance in the robustness and reliability of cold atom-based inertial sensors, especially for applications in demanding field environments.
\end{abstract}

\maketitle


Atom interferometry is of great interest for the development of inertial sensing and geodesy \cite{Peters1999,Gillot2014,Freier2016,Zhang2023,Gauguet2009, Stockton2011,Yao2021,Gautier2022}. Most precision measurements in atom interferometry have used cold atoms at few microkelvins.  However, ultracold atoms $<100\,$nK realize their full potential,  opening the possibility of large scale interferometers \cite{Plotkin2018,Gebbe2021,Beguin2023} and mitigating dispersive effects that reduce interferometer visibility \cite{Lan2012,Dickerson2013,Beaufils2023}. Ultracold atom sources, such as Bose-Einstein Condensate (BEC), often have limited production rates and typically require high power consumption. Atom chips, however, offer the advantage of creating tight magnetic traps that allow rapid production of ultracold atoms \cite{reichel2011, Rudolph2015}. In addition, they have demonstrated their compatibility with on-board experiments \cite{Becker2018, Lachmann2021}. The instability resulting from the alignment and overlapping of multiple optical beams within the vacuum chamber represents a significant challenge for the development of onboard cold atom sensor. To overcome this problem, magneto-optical (MOT) traps using optical diffraction gratings arranged on a flat surface, known as grating magneto-optical traps (GMOT), have been developed \cite{Vangeleyn2010, Lee2013, Nshii2013, Imhof2017, Sitaram2020, Bondza2022}. This technology is being explored for the miniaturization of clocks and interferometers using cold atoms at a few $\mu$K \cite{Duan2022,Lee2022,Bregazzi2024}.

In this letter, we present a hybrid device that combines the advantages of on-chip magnetic trapping with the stability of GMOT. We demonstrate the proof-of-concept of this new cold atom source by producing a rubidium-87 BEC of $6 \times 10^4$ atoms using a single laser cooling beam. The manuscript is organised such that the involved features of the chip and their performances are described following the standard ultra-cold experiment sequence. This sequence usually consists in three steps: First, the atoms are laser-cooled in a magneto-optical trap. The pre-cooled ensemble is then displaced and compressed in the vicinity of a conservative trap into which it is then transferred. Finally, the resonant light is switched off and a final evaporative cooling step allows to cross the BEC transition.
\begin{figure}
    \centering
    \includegraphics[width=0.48\textwidth]{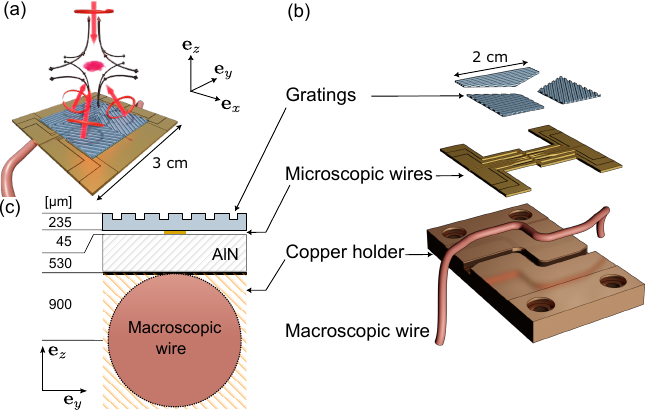}
    \caption{Atom chip assembly. 
    (a) GMOT generated by the optical gratings with a single incoming beam (red arrows), and a quadrupole field (black arrows).
    (b) Exploded-view scheme of the chip. 
    (c) Cross section of the chip including the macroscopic wire, the AlN substrate brazed with an indium layer, the microscopic wires, and the gratings.
    }
    \label{fig:principe}
\end{figure}

For the GMOT, the atom chip uses three $235\,\mu$m-thick optical gratings that are thinned and mounted in direct contact with the microscopic wires used for magnetic trapping, as shown in fig.\,\ref{fig:principe}.
To create the GMOT, a single incident laser beam illuminates all three gratings, creating a tetrahedral geometry with the $+1$ diffraction order of each grating.
In addition, the quadrupole magnetic field required for the MOT is generated by a pair of coils in anti-Helmholtz configuration outside the vacuum chamber.
The optical gratings used are described in \cite{Calviac2024}.
Their diffraction angle is $40^{\circ}$ with a Stokes parameter quantifying circular polarization of $> 90\,\%$ and a residual 0-order beam of less than 5\,$\%$. This results in a radiation pressure imbalance of less than 10\,$\%$. Detailed discussions can be found in \cite{Calviac2024, Seo2021, McGilligan2015}.
\begin{figure}
    \centering
    \includegraphics[width=0.35\textwidth]{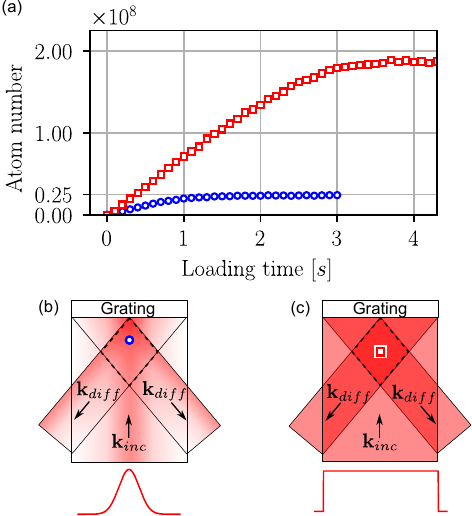}
    \caption{(a) GMOT loading with an incident Gaussian beam (waist 22.5\,mm, blue circles) and with a flat top beam (radius 10.5\,mm, red squares) with the same total power of $56$\,mW.
    (b) Schematic view of the beam overlap in the MOT region for a Gaussian incident beam.
    (c) Same for a flat top incident beam.
    The blue dot (resp. red square) represents the center-of-mass position of the loaded cloud for both configurations located respectively 3.5\,mm and 5.7\,mm away from the surface.
    }
    \label{fig:gmot}
\end{figure}

The GMOT is loaded from a 2D-MOT using an incident beam power of 56\,mW red-detuned by $\Delta=-1.8\,\Gamma$ from the cooling transition where $\Gamma = 2\pi\times6.07$\,MHz is the linewidth of the $^{87}$Rb D2 transition ($5^2S_{1/2} - 5^2P_{3/2} $).
Figure\,\ref{fig:gmot}\,(a) shows the atom number loaded in the GMOT for a Gaussian beam with a waist of 22.5\,mm and a flat top beam with a radius of 10.5\,mm, achieving respectively $2.5 \times 10^7$ and $2 \times 10^8$ atoms.
This improvement is due to a more uniform optical power across the overlapping beams with the flat top beam, along with a trap center positioned farther from the surface.
In particular, the inhomogeneous illumination of the Gaussian beam imposes to displace the quadrupole field on the edge of the overlap volume (3.5 mm away from the grating surface) \cite{Seo2021}. In contrast, the uniform illumination of the flat top beam allows utilizing the entire volume by positioning the MOT at 5.7\,mm from the surface (see fig.\,\ref{fig:gmot}\,(b-c)).
Therefore, this configuration results in reduced losses, as evidenced by the trap lifetime increased from 2.0\,s with the Gaussian beam to 10.5\,s with the flat top beam, and an increased capture volume \cite{McGehee2021,HendrikThese2023}. In addition, the flat top beam greatly improves the GMOT's robustness to beam misalignment. All the subsequent experiments have been performed with the flat top configuration.

\begin{figure*}
    \centering
    \includegraphics[width=0.7\textwidth]{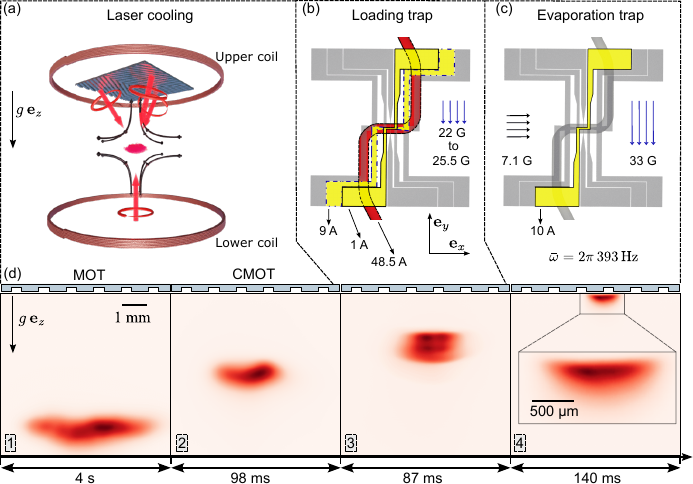}
    \caption{Loading of the magnetic traps.
    (a) Scheme of the Laser cooling configuration where the magnetic quadrupole field coils are represented. 
    (b) Front view of the magnetic chip structure where the wires used at the loading stage are colored. The loading trap uses the macroscopic wire (red), the 6\,mm (dash yellow) and the 2\,mm-long (solid yellow) microscopic wires with the indicated currents. The bias field (blue arrows) along the $y$-axis is ramped from 22\,G to 25.5\,G during the loading.
    (c) Same for the magnetic trap used for the evaporation step where only the 2-mm-long wire is used.
    A second external field along the $x$-axis (black arrows) decreases the magnetic field minimum to 0.8\,G.
    (d) Fluorescence images of the cloud at different steps of the loading process : (1) MOT, (2) before magnetic loading, (3) in the loading trap, (4) in the evaporation trap. The duration of each step is indicated below the images.
    }
    \label{fig:magtrap}
\end{figure*}

To achieve an efficient loading of the laser-cooled atoms in the magnetic trap, it is essential to increase the phase-space density of the atomic ensemble and to move the atoms closer to the center of the magnetic trap located at most 3.1\,mm away from the surface. However, due to the size of the cloud and the conical shape of the beam overlap region, loading the GMOT directly around the magnetic trap position is not feasible without significant loss. Therefore, a compromise is made by moving the cloud from where the loading is maximized to within 3.5\,mm of the surface. 
In practice, losses are minimized by a compressed MOT step performed by simultaneously reducing the laser power to 33\,mW and increasing the detuning to -3.7$\,\Gamma$, while increasing the current only in the lower magnetic coil (see fig.\,\ref{fig:magtrap}\,(a)). The temperature is then further reduced by linearly increasing the detuning to -14.7$\,\Gamma$. Simultaneously, the currents in the two coils generating the quadrupole field are ramped down. The current in the upper coils cancels out while the one in the lower coil is kept non-zero. The non-symmetric extinction of the quadrupole field coils controls the final displacement of the cloud towards the chip surface. 
The shape of the atomic ensemble after these two first laser cooling steps are given in the two first panels of fig.\,\ref{fig:magtrap}\,(d).
Finally, we optically pump the $8\times 10^7$ atoms at $35\,\mu$K in the weak field-seeking state with the maximal magnetic moment $\vert F=2, m_F = +2\rangle$. 

In our setup, the atoms are trapped in magnetic traps produced by planar Z-shaped wires with different central segment lengths arranged on two layers \cite{reichel2011}. A first layer consists of $45\,\mu$m-thick and $300\,\mu$m-wide copper wires electrodeposited on an AlN substrate (see fig.\,\ref{fig:principe}\,(c)).
The relatively large wire section is a key ingredient to minimize the Joule heating. It is designed considering its influence on the trap depth and the transverse gradients at a distance compatible with the thickness of the optical grating \cite{Squires2011}.
These wires can form Z-traps with segment lengths of 10\,mm, 6\,mm, or 2\,mm, which form the final magnetic trap where the BEC is obtained. Note that the presented sequence only use the two shorter wires (colored in yellow in fig.~\ref{fig:magtrap}\,(b-c)). Beneath the layer of microscopic wires, a secondary layer contains a 10\,mm Z-trap with large-diameter wires (see fig.~\ref{fig:principe}\,(c)) to accommodate the higher currents required for efficient atom transfer from the GMOT to the magnetic trap. To improve heat transfer, the AlN substrate is indium soldered onto the copper mount.

The loading trap is created by the combination of a bias field of 22\,G and the field generated by the macroscopic wire (48.5\,A) \cite{Alibert2017}. In addition, smaller currents (9\,A and 1\,A respectively) are sent through the two microscopic wires of length 6\,mm and 2\,mm as shown in fig.\,\ref{fig:magtrap}\,(b). This configuration captures most of the laser cooled atoms at a distance of 3.1\,mm from the chip surface. The bias field is then further increased to 25.5\,G to displace the trapped cloud even closer to the surface (2.3\,mm) (third panel of fig.\,\ref{fig:magtrap}\,(d)).

The loading trap does not provide high enough trapping depth and frequencies to ensure an efficient evaporative cooling \cite{Reichel2002}. Therefore, the cloud is transferred to a steeper trap: the evaporation trap.
This trap is created by a stronger bias field (33\,G) and 10\,A sent through the 2-mm-long wire.
A second homogeneous external field is applied along the wire direction to reduce the value of the magnetic field minimum and to increase the longitudinal stiffness of the trap (see fig.\,\ref{fig:magtrap}\,(c)).
The magnetic field minimum is located at $\mathtt{\sim}\,265\,\mu$m from the surface, and the trap depth is limited by the gravitational sag to $1.7$\,mK.
We characterize the trap by measuring the frequencies of low-amplitude dipole oscillations induced by sudden changes in the wire current.
For the evaporation trap, this analysis yields $1014 \times 1014 \times 59$\,Hz$^3$, resulting in an average trapping frequency of 393\,Hz, in agreement with numerical simulations.

In practice, the different steps of the laser cooling and the loading of the magnetic trap have been optimized with a Bayesian optimization strategy \cite{wigley2016,barker2020,Yu2024} using the total atom number or the central density of the cloud as cost functions. Finally, $2\times 10^7$ atoms, representing around 10\,\% of the atoms initially loaded in the MOT, are trapped in the evaporation trap at a temperature of the order of $150\,\mu$K.

\begin{figure}
    \centering
    \includegraphics[width=0.35\textwidth]{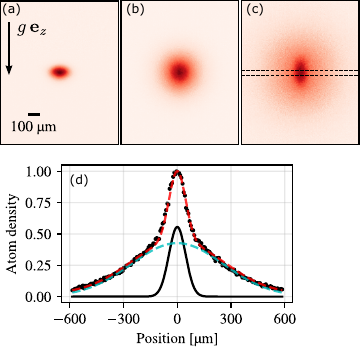}
    \caption{(a-b-c) Fluorescence images after 4.5\,ms, 20.5\,ms and 44.5\,ms of time-of-flight from the loose trap of an evaporatively cooled ensemble slightly below the critical temperature. 
    (d) Vertical sum of the image in panel (c) in between the two dashed lines. The bimodal distribution is fitted by a sum of two Gaussians (red dashed line). The solid black line (resp. dashed blue line) shows the individual fit of the condensed (resp. thermal) component. 
    }
    \label{fig:BEC}
\end{figure}

The final cooling step is a forced evaporation cooling induced by Radio-Frequency (RF) transition to untrappable states. The RF current is fed to the macroscopic wire and its frequency is ramped from 25\,MHz to 0.54\,MHz in 3.2\,s following $4$ linear segments. The ramp parameters - duration, start and stop frequencies - were optimized using the same Bayesian algorithm. The Bose-Einstein transition occurs at a temperature of 350\,nK with $6 \times 10^4$ atoms. By pushing the evaporation to lower temperature we can achieve a fully condensed ensemble (with no detectable thermal component) with $3.5\times 10^4$ atoms.

Fluorescence imaging of the atomic cloud is performed after a sudden trap release. The final trap is very tight and very close to the surface, so there are significant atomic losses due to collisions with the surface. To obtain accurate quantitative estimates of atom number and temperature from time-of-flight measurements, the evaporatively cooled atoms are smoothly transferred to a shallower trap, the loose trap. This transfer is achieved by linearly ramping the two external magnetic fields over 50\,ms. The bias field along the $y$-axis is decreased from 33\,G to 16\,G and the external field along the $x$-axis increased from 7.2\,G to 9\,G in order to keep the minimum field around 1\,G.
The loose trap has a maximum depth of $669\,\mu$K at a distance of $\mathtt{\sim} 665\,\mu$m from the surface and an average trapping frequency of 202\,Hz ($437\times437\times43$\,Hz$^3$).
   
Figure\,\ref{fig:BEC}\,(a-b-c) shows time-of-flight images of a partially condensed cloud for short, intermediate, and long free-fall times out of the loose trap. At short times, the distribution reflects the trap aspect ratio showing an elongated cloud along the weak trapping direction ($x$-axis). At intermediate times, the distribution becomes isotropic. At long times, the distribution presents the typical bimodal behavior with an isotropic expansion of the thermal component and an anisotropic expansion of the condensed component whose interaction energy is mostly released in the tightly confining direction ($z$-axis). This well-known ellipticity inversion is a signature of Bose-Einstein condensation in an anisotropic trap \cite{Castin1996}. The bimodal distribution is highlighted in fig.\,\ref{fig:BEC}\,(d). It presents a vertical sum of a small central horizontal strip of the distribution after long time-of-flight, fitted by a two Gaussians model. 
    
Figure\,5 shows a quantitative study of the crossing of the condensation transition in the loose trap. It shows how the condensed fraction evolved when decreasing the temperature around the critical point.
The atom number in both thermal and condensed components is extracted by double Gaussian fits to the entire cloud, similar to fig.\,\ref{fig:BEC}\,(d).
Knowing the total number of atoms and the average trapping frequency, $\bar \omega / (2 \pi)$, of the loose trap, we calculate the critical temperature as $T_c =\left(\frac{N}{\zeta(3)}\right) ^{\frac{1}{3}} \hbar \bar{\omega}/k_\mathrm{B}$, with $\zeta(3)\sim1.202$ \cite{Dalfovo1999}. The temperature $T$ of the cloud is measured by a standard ballistic expansion of the thermal part of the cloud during time-of-flight. The data are compared to the expected evolution in $1-(T/T_c)^3$ of the normalized condensed fraction. The appearance of a detectable condensed fraction and its growth with decreasing temperature are consistent with the independent measurements of the atom number and the trapping frequency. All the evidences presented above confirm that our hybrid atom chip offers the required features to produce ultra-cold atoms on a robust and reliable apparatus.

\begin{figure}
    \centering
    \includegraphics[width=0.45\textwidth]{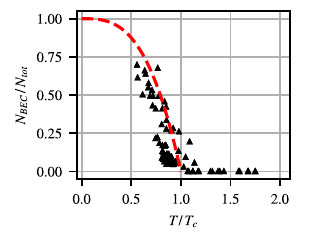}
    \caption{Evolution of the condensed fraction around the critical temperature.
    The red dashed line shows the expected $1-(T/T_c)^3$ behavior. 
    }
    \label{fig:TC}
\end{figure} 

In summary, we have successfully demonstrated the first BEC source combining optical gratings and a magnetic trap on a single chip. This approach allowed the realization of a Rubidium-87 BEC with about $6\times 10^4$ atoms using a single laser cooling beam. The BEC is achieved after a cooling sequence of $\sim8$\,s. The production rate reported here, one order of magnitude below state-of-the-art with atom chip \cite{Rudolph2015}, can be improved with our current chip design.
For example, \cite{HendrikThese2023} reported an order of magnitude larger number of atoms in the GMOT using a similar apparatus. In addition, using the multiple unused wires present on our chip to create steeped traps has the potential to accelerate evaporative cooling, thereby increasing BEC production rates. These advances suggest the possibility of developing highly robust ultracold atom sources, particularly suitable for on-board applications. In addition, this technology paves the way for novel optical architectures in quantum technologies, broadening the range of applications. For example, diffraction of Raman beams on optical gratings would allow acceleration measurements in all three spatial directions using different combinations of diffracted wave vectors, paving the way for 3D inertial sensors \cite{Barrett2019}.

\vspace{0.5cm}
\begin{acknowledgments} This work was supported by ANR  funding grants MiniXQuanta ANR-20-CE47-0008, QAFCA ``Plan France 2030'' ANR-22-PETQ-0005, and EUR grant NanoX ``Programme des Investissements d’Avenir'' ANR-17-EURE-0009, and by CNES under action R\&T R-S20/SU-0001-036. RC acknowledges support from CNES and Region Occitanie. This work was supported by LAAS-CNRS micro and nanotechnologies platform member of the French RENATECH network.
\end{acknowledgments}

\end{document}